\documentclass[twocolumn,showpacs,preprintnumbers,amsmath,amssymb,prl,superscriptaddress]{revtex4}

\usepackage{graphicx}
\usepackage{dcolumn}
\usepackage{bm}

\def\adot{\dot{\alpha}}

\begin{document}


\title{Limit on the Temporal Variation of the Fine-Structure Constant\\ Using Atomic Dysprosium}

\author{A. Cing\"{o}z}
\author{A. Lapierre}
 \affiliation{Department of Physics, University of California at
Berkeley, Berkeley, California 94720-7300, USA}%
\author{A.-T. Nguyen}
\affiliation{Los Alamos National Laboratory, Physics Division, P-23,
MS-H803, Los Alamos, New Mexico 87545, USA}
\author{N. Leefer}
\affiliation{Department of Physics, University of California at
Berkeley, Berkeley, California 94720-7300, USA}
\author{D. Budker}
\affiliation{Department of Physics, University of California at
Berkeley, Berkeley, California 94720-7300, USA}%
\affiliation{Nuclear Science Division, Lawrence Berkeley National
Laboratory, Berkeley, California 94720, USA}%
\author{S. K. Lamoreaux}
\altaffiliation[Present address: ]{Department of Physics, Yale
University, New Haven, Connecticut 06520-8120, USA}
 \affiliation{Los Alamos National Laboratory, Physics Division, P-23,
MS-H803, Los Alamos, New Mexico 87545, USA}
\author{J. R. Torgerson}
\affiliation{Los Alamos National Laboratory, Physics Division, P-23,
MS-H803, Los Alamos, New Mexico 87545, USA}

\date{\today}
\begin{abstract}
Over a period of eight months, we have monitored transition
frequencies between nearly degenerate, opposite-parity levels in two
isotopes of atomic dysprosium (Dy). These transition frequencies are
highly sensitive to temporal variation of the fine-structure
constant ($\alpha$) due to relativistic corrections of large and
opposite sign for the opposite-parity levels. In this unique system,
in contrast to atomic-clock comparisons, the difference of the
electronic energies of the opposite-parity levels can be monitored
directly utilizing a radio-frequency (rf) electric-dipole transition
between them. Our measurements show that the frequency variation of
the 3.1-MHz transition in $^{163}$Dy and the 235-MHz transition in
$^{162}$Dy are 9.0$\pm$6.7~Hz/yr and -0.6$\pm$6.5~Hz/yr,
respectively. These results provide a value for the rate of
fractional variation of $\alpha$ of $(-2.7\pm2.6)\times
10^{-15}$~yr$^{-1}$ (1 $\sigma$) without any assumptions on the
constancy of other fundamental constants, indicating absence of
significant variation at the present level of sensitivity.
\end{abstract}

\maketitle

Modern theories attempting to unify gravitation with the other
fundamental interactions have renewed interest in experimental
searches for temporal and spatial variation of fundamental
constants. These theories allow for, or even predict, such
variation, which would violate Einstein's Equivalence Principle
~\cite{Uzan2003}. Recently, evidence for variation of $\alpha$ over
cosmological time scales was discovered in absorption spectra of
light from quasars~\cite{Webb2001, Murphy2003}. The result in
Ref.~\cite{Murphy2003} corresponds to
$\adot/\alpha=(6.40\pm1.35)\times10^{-16}$/yr assuming a linear
shift over $10^{10}$ years. However, more recent measurements
present sensitive results that are consistent with no
variation~\cite{Quast2004,Srianand2004}. On a geological time scale
of $10^9$ years, a test for variation of $\alpha$ comes from
analyses of fission products of a natural reactor in Oklo (Gabon).
There is a discrepancy between earlier
analyses~\cite{Damour1996,Fujii2000}, which are consistent with no
variation, and the analysis in Ref.~\cite{Lamoreaux2004}, where a
variation was reported. However, a recent full Monte Carlo
simulation of the reactor~\cite{Gould2006} reports a result
consistent with no variation at the level of $1.2\times10^{-17}$/yr.

Observational measurements are difficult to interpret due to
numerous assumptions and uncontrollable systematic uncertainties.
Laboratory searches (see, for example, Refs.~\cite{Bize2003,
Marion2003, Fischer2004, Peik2004}), which probe variations on the
time scale of years, are easier to interpret since they are
repeatable, and systematic uncertainties can be studied by changing
experimental conditions. The best laboratory limit of
$|\adot/\alpha|<1.2\times10^{-15}$/yr was obtained from a comparison
of a Hg$^+$ optical clock to a Cs frequency
standard~\cite{Bize2003}. This result assumes that other constants,
in particular $m_e/m_p$, do not vary in time~\cite{note:memp}.
Interestingly, evidence for variation of this parameter on
astronomical time scale was recently reported in
Ref.~\cite{Reinhold2006}. The best limit that is independent of
assumptions regarding other constants was obtained by combining this
result with a comparison of Yb$^+$ optical clock and a Cs frequency
standard~\cite{Peik2004}. This measurement gives
$|\adot/\alpha|<2.2\times10^{-15}$/yr.

In this Letter, we report a limit on the temporal variation of
$\alpha$ that is independent of other constants. This result is
based on observing, over a period of eight months, rf-transitions
between nearly degenerate, opposite-parity levels in atomic
dysprosium, as suggested in Ref.~\cite{Dzuba1999b}.

\begin{figure}
\includegraphics{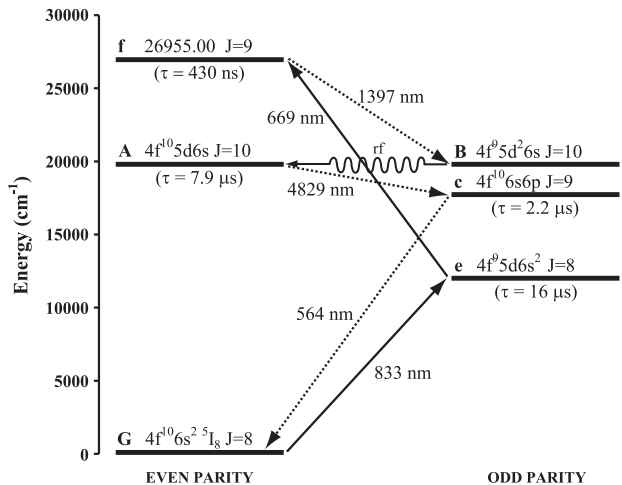}
\caption{\label{fig:popsch} Relevant levels and transitions in
atomic dysprosium used to populate level $B$ and detect the
population of level $A$. Solid lines indicate laser excitation;
dashed lines indicate spontaneous emission; wavy line indicates rf
electric field.}
\end{figure}

The energy of an atomic level can be expressed as
\begin{equation}
E=E_0+q(\alpha^2/\alpha_0^2-1),
\end{equation}
where $E_0$ is the present-day energy, $\alpha_0$ is the present-day
value of the fine-structure constant, and $q$ is a level dependent
coefficient that determines the sensitivity to variations in
$\alpha$~\cite{Dzuba2003}. A theoretical calculation has found that
the $q$-coefficients for the two nearly degenerate, opposite-parity
levels in Dy (Fig.~\ref{fig:popsch}, levels $A$ and $B$) are large
and of opposite sign~\cite{Dzuba2003}. For the even-parity level
(level $A$), $q_A/hc\simeq6\times10^3$~cm$^{-1}$, while for the
odd-parity level (level $B$), $q_B/hc\simeq-24\times10^3$~cm$^{-1}$.
Since explicit uncertainties are not quoted for these numbers, they
are assumed to be exact in the determination of the
result~\cite{note:qunc}.

Another parametrization of frequency and sensitivity factor is often
used for atomic-clock comparisons, in which an electronic transition
frequency is expressed as
\begin{equation}
\nu=R_\infty A F(\alpha),
\end{equation}
where $R_\infty$ is the Rydberg constant expressed in units of
frequency, $A$ is a dimensionless atomic structure factor
independent of $\alpha$, and $F(\alpha)$ is a factor that includes all relativistic as well as many-body effects which depend on
$\alpha$~\cite{Prestage1995}. The change in a transition frequency can be related to a fractional change
in $\alpha$ as
\begin{equation}
\frac{\dot{\nu}}{\nu}=\frac{d}{dt}\ln \nu=\left[\alpha\frac{\partial \ln F}{\partial\alpha}\right]\frac{\adot}{\alpha}.
\end{equation}
The expression $\alpha~\partial\ln F/\partial\alpha$ is the
sensitivity of a transition to variation in $\alpha$. Using this
parametrization, the q-coefficients for atomic Dy can be recast (for
$\alpha\approx\alpha_0$) to give $\alpha~\partial\ln
F/\partial\alpha = 2q/(h\nu)$, where $h\nu$ is the energy of level
$A$ or $B$ with respect to the ground state. This gives a
sensitivity of 0.6 and $-2.4$ for level $A$ and $B$, respectively.
These values are comparable, for example,  to the sensitivity of
-3.2 for the transition used in the Hg$^+$ optical
clock~\cite{Dzuba1999}.

In atomic-clock comparisons, the observable quantity is the ratio of
the two frequencies being compared. A unique aspect of the Dy system
is that the direct observable quantity is the difference
$\nu_B-\nu_A$, due to the fact that an electric-dipole transition
can be induced between the two levels using a rf electric field. The
time variation of the transition frequency between levels $A$ and
$B$ is
\begin{equation}
\label{eq:sens}
\dot{\Delta\nu}=\dot{\nu}_B-\dot{\nu}_A=2\frac{q_B-q_A}{h}\frac{\adot}{\alpha}\sim-1.8\times10^{15}
\mbox{Hz} \frac{\adot}{\alpha}.
\end{equation}
For instance, $|\adot/\alpha|=10^{-15}$/yr implies $|\dot{\Delta
\nu}|\simeq2$~Hz/yr.

There are several advantages to using the nearly degenerate levels
in Dy. Because most of the transition frequencies are $\sim$1~GHz or
smaller, direct frequency counting techniques can be used. This
allows for the comparison of two electronic transitions without the
need for optical frequency combs or transfer cavities. A preliminary
analysis of statistical and systematic uncertainties shows that the
measurement of the transition frequency and the control of the
systematics at a mHz level is feasible, which corresponds to an
ultimate sensitivity of $|\adot/\alpha|\sim10^{-18}$/yr for two
measurements separated by a year's time~\cite{Nguyen2004}. A mHz
resolution on a transition frequency of 1~GHz requires a relatively
modest fractional stability of $10^{-12}$ for the reference
frequency standard. This also means that the results are insensitive
to variation of the Cs reference frequency due to changes in the
values of fundamental constants since the experimental upper limit
on this relative variation rate is $\sim 10^{-15}$~yr$^{-1}$ (see,
for example,~\cite{Marion2003}). In addition, because the isotope
shifts and hyperfine splittings are on the order of the transition
frequencies, it is possible to work with multiple rf transitions
corresponding to the same electronic transition, including, in
particular, transitions with energy differences of opposite sign.
Since the variation in $\alpha$ only depends upon electronic energy
levels~\cite{note:hyperfine}, the variation of these rf transitions
should have equal magnitude but opposite sign. This correlation can
be used to detect and eliminate certain systematics. Currently, we
monitor two such transitions: the 3.1-MHz ($F=10.5\rightarrow
F=10.5$) transition in $^{163}$Dy and the 235-MHz ($J=10\rightarrow
J=10$) transition in $^{162}$Dy.

The population and detection scheme is shown in
Fig.~\ref{fig:popsch}. The long-lived ($\tau_B
>200$~$\mu$s~\cite{Budker1994}) level $B$ is populated by
three transitions. The first two transitions are driven with 833-
and 669-nm laser light, while the third transition involves
spontaneous decay. Atoms are transferred to level $A$ with rf
electric field referenced to a Cs frequency standard. Level $A$
decays to the ground state in two steps. Fluorescent light at 564 nm
from the second decay is used for detection.

The atomic beam is described in Ref.~\cite{Nguyen1997}. It is
produced by an effusive oven operating at $\sim$1500~K. In addition
to a multislit nozzle-array attached to the oven, two external
collimators are used to collimate both the atomic beam and the oven
light (in order to minimize the background due to scattered light
from the oven). The resultant atomic beam has a mean velocity of
500~m/s and a full-angle divergence of $\sim$0.2~rad
(1/$e^2$-level).

The laser light at 833 and 669 nm is produced by a Ti:Sapphire ring
laser and a ring dye laser with DCM dye, respectively. Since
narrow-band continuous-wave lasers are inefficient at exciting
atomic beams with weak collimation, an adiabatic passage technique
is utilized to transfer the population to level $f$ (a description
of this technique, as well as references to earlier work, are given
in Ref.~\cite{Nguyen2000}). Briefly, the beams are diverged with
cylindrical lenses to match the atomic-beam divergence. Due to the
Doppler effect, the atoms experience a frequency chirp in laser
detuning which transfers the population to the excited state with
high efficiency. The laser-light and rf-interaction region is
enclosed by a magnetic shield. The residual field in the interaction
region is $\sim$1~mG.

The rf-generation and detection system is discussed in
Ref.~\cite{Cingoz2005}. The frequency modulated rf field is
generated by a synthesizer referenced to a commercial Cs frequency
standard which is compared to a second clock (Rb oscillator locked
to a digital cellular network signal) to monitor its stability. The
short term fractional stability of the clocks is $\sim10^{-11}$, and
it approaches $10^{-12}$ in 10 minutes of integration. The
modulation frequency is 10~kHz with a modulation index of 1. Since
the rf-transition linewidth is $\sim$20~kHz, this modulation
provides a fast sweep across the line shape, minimizing the effect
of fluctuations such as those due to laser power drifts or density
fluctuations of the atomic beam. The modulation is provided by the
reference output of a lock-in amplifier which demodulates the signal
from a photomultiplier tube used to detect the fluorescence. The
line shape of the first- and second-harmonic outputs of the lock-in
amplifier is, respectively, an odd function with zero crossing on
resonance and an even function with a maximum on resonance. In order
to reduce drifts further, the ratio of these two is used to measure
the transition frequency, which is extracted by a two-step process
described in detail in Ref.~\cite{Cingoz2005}.

There are several imperfections that affect the stability of our
transition-frequency measurements. The largest effect is due to the
combination of the residual magnetic field with the laser-light
polarization imperfections. During normal operation, the
polarizations of both laser beams are set to linear, which should
lead to symmetrically populated magnetic sublevels. However,
residual light ellipticity leads to atomic orientation, which, in
the presence of the residual magnetic field, leads to asymmetric
broadening of the line and causes apparent shifts~\cite{Nguyen2004}.
In systematic studies where we deliberately used circularly
polarized light, shifts in the transition frequencies as large as
$\pm$160~Hz were observed. The linear polarizations of the laser
beams are determined by calcite polarizers with $10^{-5}$ extinction
ratio outside the vacuum chamber. However, the beams must travel
through optical windows and lenses utilized for adiabatic passage
population technique (see above) before reaching the interaction
region. A small amount of stress-induced birefringence on these
lenses was discovered by deliberate misalignment of the laser beams.
The amount of ellipticity depended upon where the beam sampled a
lens. The residual systematic uncertainties due to this effect range
from 2 to 5~Hz on different experimental runs with the smaller
uncertainty corresponding to later runs in which the lens mounts
were modified to relieve the stress.

In addition to the residual magnetic field, the magnetic shielding
was inadvertently magnetized in the April 2006 runs during studies
utilizing external magnetic-field coils. Since the amount of
magnetization and the resultant shifts in the transition frequencies
were unknown at the time, systematic uncertainties have been
assigned by driving the magnetization to saturation and noting the
maximum shift induced for both field polarities. The uncertainty due
to this effect is 7.5~Hz. The shields were demagnetized at the
beginning of each subsequent run.

Another important systematic uncertainty is due to the collisions of
Dy atoms with the background gas in the vacuum chamber. Collisional
shift rates for the rf transitions due to various gases were
determined in Ref.~\cite{Cingoz2005}. The pressure in the vacuum
chamber is $\sim 5\times 10^{-7}$~Torr when the Dy oven is off. When
the oven is turned on, the pressure rises to $\sim 2\times
10^{-6}$~Torr, limited by H$_2$ outgassing from the oven. In
Ref.~\cite{Cingoz2005}, the observed shift rates for H$_2$ were
found to be consistent with zero. However, the pressure of other
gases, such as N$_2$, also increases at the level of $\sim 4\times
10^{-7}$~Torr when the oven is turned on, and continually decreases
and stabilizes after $\sim 6$~hours of continuous oven operation.
The partial pressures are monitored with a residual gas analyzer
(RGA), and the transition frequencies are corrected for the presence
of N$_2$, O$_2$, H$_2$, and Ar using the shift rates measured for
each gas. The variation in the pressure of other gases such as
H$_2$O and CO$_2$ are $\sim10^{-7}$~Torr or smaller, and no
significant correlation between these pressure variations and the
transition frequency is observed. The total correction is $<2$~Hz at
any time during each run. The results also include small
uncertainties ($<0.6$~Hz) due to oven temperature variations, which
may indicate a possible effect of Dy intra-beam
collisions~\cite{Cingoz2005}.

Finally, there are systematic effects associated with rf
electric-field inhomogeneities. The rf electrodes are fed by
twisted-pair wires at one corner. Numerical simulations show that
with this feed, at rf wavelengths comparable to the size of the
electrodes ($\sim 10$~cm), spatially varying amplitude and phase
inhomogeneities become substantial. These inhomogeneities can lead
to drifts in the transition frequency when they are combined with
changes in the atomic beam velocity distribution. This effect was
studied experimentally by deliberate detuning of the laser
frequencies from the optical resonances. This leads to changes in
the transverse velocity distribution of the excited atoms due to
imperfect adiabatic passage. For the two transitions considered in
this study, shifts of $<0.5$~Hz for 1~MHz detuning in the laser
frequencies were observed. To keep this effect at the 1~Hz level
during the runs, the laser frequencies were periodically retuned
($\sim$ every 2 minutes) to the center of their respective
transitions. To check for slow variations in the beam velocity
distribution due to other changes such as collimator clogging, this
study was repeated several times over a two-year period, during
which the oven was reloaded and collimators cleaned several times.
No significant change in the size or dependence of the effect was
observed.

All other systematics considered in Ref.~\cite{Nguyen2004} were
estimated to be smaller than 1 Hz for the specific conditions of the
current experiment. Many of these systematics, such as AC Stark
shifts and oven black-body radiation shifts, were studied
experimentally by exaggerating certain imperfections and were found
to be negligible.

\begin{figure}
\includegraphics{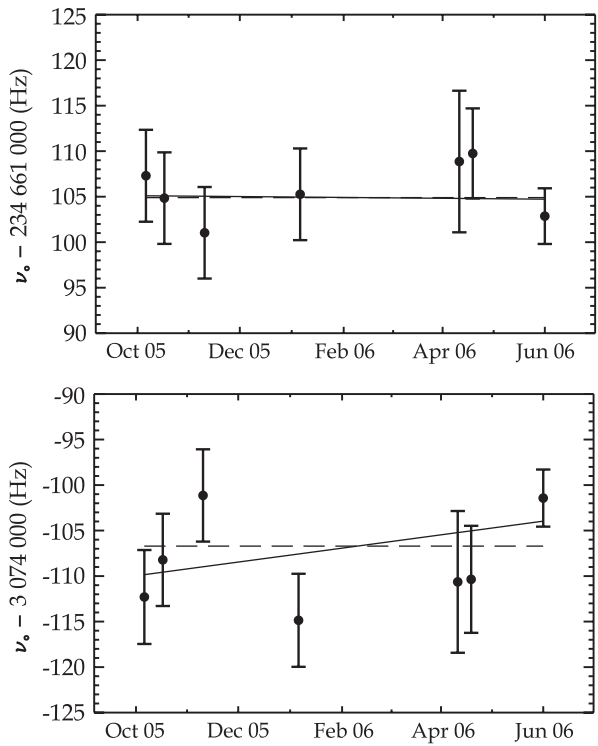}
\caption{\label{fig:result} Measured transition frequencies for the
235-MHz and 3.1-MHz transitions over an eight-month period. The data
have been corrected for collisional shifts. The solid lines are the
least-squares linear fit to the data. The dashed lines are the
least-squares fit to a constant function.}
\end{figure}

We have measured the two rf transition frequencies in the course of
seven runs over eight months. Figure~\ref{fig:result} shows the
results of these measurements corrected for collisional shifts.
Uncertainties are due both to statistical and aforementioned
systematic uncertainties. Least-squares linear fit to the data
points gives slopes of $-0.6\pm 6.5$~Hz/yr and $9.0\pm6.7$~Hz/yr for
the 235-MHz and 3.1-MHz transitions, respectively. According to Eq.
(\ref{eq:sens}), these results correspond to
$\adot/\alpha=(-0.3\pm3.6)\times 10^{-15}$~yr$^{-1}$ for the 235-MHz
transition and $(-5.0\pm3.7)\times 10^{-15}$~yr$^{-1}$ for the
3.1-MHz transition. As mentioned earlier, because the energy
differences for these two transitions are of opposite sign, these
two data sets can be combined to construct sum and difference
frequency plots. The sum frequency slope, which should be
insensitive to a variation in $\alpha$, is $8.1\pm9.4$~Hz/yr and
consistent with zero. The difference frequency, which is twice as
sensitive to a variation in $\alpha$ as the individual frequencies,
gives the final result of
$\adot/\alpha=(-2.7\pm2.6)\times10^{-15}$~yr$^{-1}$, consistent
(1~$\sigma$) with no variation at the present level of sensitivity.

In conclusion, we have presented the first result of a direct
measurement of the temporal variation of $\alpha$ with atomic
dysprosium. This result is of comparable uncertainty to the present
best laboratory result that is independent of other fundamental
constants. However, in our case the interpretation does not require
comparison with different measurements to eliminate dependence on
other constants, and provides an alternative to measurements that
utilize state-of-the-art atomic optical frequency clocks. The
present uncertainty is dominated by systematic effects primarily due
to polarization imperfections coupled to the residual magnetic
field, collisional shifts, and rf electric-field inhomogeneities. A
significant improvement is expected from a new apparatus under
construction which will provide better control over these, as well
as other, systematic effects expected to be important to achieve
better than 1-Hz sensitivity. Ultimately, mHz-level sensitivity may
be achievable with this method~\cite{Nguyen2004}.

We thank V. V. Yashchuk, D. F. Jackson Kimball, and D. English for
valuable discussions. This work was supported in part by the
University of California - Los Alamos National Laboratory CLC
program, NIST Precision Measurement Grant, Los Alamos National
Laboratory LDRD, and by grant RFP1-06-15 from the Foundational
Questions Institute (fqxi.org).

\bibliography{prlpaper,footnotes}

\begin{thebibliography}{26}
\bibitem[{not({\natexlab{c}})}]{note:hyperfine}
\bibinfo{note}{There are differences due to the hyperfine interaction and
  isotope shifts, but these are several orders of magnitude smaller.}
\end{thebibliography}
\end{document}